\documentstyle[12pt]{article}
\begin{document}
\newfont{\elevenmib}{cmmib10 scaled\magstep1}%
\newfont{\cmssbx}{cmssbx10 scaled\magstep3}
\newcommand{\preprint}{
            \begin{flushleft}
            \elevenmib Northwest\, University\\
            \elevenmib Graduate\, School\, of\, Academy \, Kyoto
            \end{flushleft}\vspace{-1.3cm}
            \end{flushright}}
\newcommand{\Title}[1]{{\baselineskip=26pt \begin{center}
            \Large   \bf #1 \\ \ \\ \end{center}}}
\newcommand{\Author}{\begin{center}\large \bf
           Bo-yu Hou$^a$\footnote[1]{e-mail:\ 
                 {\tt byhou@phy.nwu.edu.cn}  },  
           Bo-yuan Hou$^b$\footnote[2]{e-mail:\
                 {\tt byuanhou@phy.nwu.edu.cn}  } and 
            Ruihong Yue$^a$\footnote[3]{e-mail:\
                   {\tt yue@phy.nwu.edu.cn}%
          }     \end{center}}
\newcommand{\Address}{\begin{center} \it
            $^a$Institute of Modern Physics, Northwest University,
            Xi'an 710069, P. R. China \\
            $^b$Graduate School, Chinese Academy of Science, 
             Beijing 100039, P. R. China \end{center}  }
\baselineskip=20pt

\thispagestyle{empty}
\bigskip
\bigskip
\bigskip
\Title{ Fuzzy sphere bimodule, ABS construction to the exact soliton
solution}
\Author

\Address
\vspace{2cm}

\begin{abstract}%
\noindent
In this paper, we set up the bi-module of the algebra ${\cal A}$ on
fuzzy sphere. Based on the differential operators in moving frame, we
 generalize the ABS construction into fuzzy sphere case. The applications
of ABS construction are investigated in various physical systems.
\\   \\
PACS Numbers: 46.10.+z, 05.40.+j, 05.60.+w
\end{abstract}

\section{Introduction}

Non-commutative geometry is originally an old topics in
mathematics \cite{conn1}. Now, it  becomes an  interesting subject of
quantum
field theory since it was  found that the
non-commutative field theory appears naturally in string theory at
low energy level in a constant NS B-field \cite{cds,sw}. The nontrivial
B-field
leads to the non-commutativity of the coordinates of string ends
on the D-brane, which gives a non-commutative gauge field theory
on the D-brane world volume.

  Recently, the study of the non-perturbative dynamics of these
fields has attracted much more attention [4-20]. Harvey, Kraus and 
Larsen set up a new method to investigate the soliton solution , 
the monopole solution and instanton solution in 3+1
dimension \cite{hkl}. The Nielson-Olesen solution in abelian Higgs model
was
studied in \cite{jmw} and \cite{bak}. All studies are carried
out in the non-commutative Euclidean space. One natural question
is to how to generalize such results into other non-commutative
geometry with non-zero curvature. It is a challenging topics.
Physically, the space near horizon in NS 5-brane is a sphere
$S^3$. The D-brane on such geometry can be described  by a
boundary WZW model on a sphere $S^2$ \cite{ars}.  Many works
focused on the fuzzy sphere $S^2$-- a simplest space with
non-trivial curvature \cite{gkp}. On the fuzzy sphere, one can set up
a gauge filed theory \cite{ww} directly or from the matrix theory
\cite{kimura,iktw}. The D-brane structure was discussed in
\cite{hnt,hkl}. The $CP(N)$ system on fuzzy sphere in ref.
\cite{ccly}.

  In the present paper, we will consider the gauge field theory on
fuzzy sphere, which includes the soliton and Nielsen-Olesen
solution. The main tool  is  ABS construction. In the standard ABS
method, the quasi-unitarity operator has played an important role.
However, on the fuzzy sphere, there does not exist such operator.
We will generalize it and find a partial isometric operator which
acts a same role as quasi-unitarity operator in usual one.

 On the sphere, the spherical harmonic functions are convenient basis for
studying other properties of the theory. But, most study about
fuzzy sphere is based on the three-dimensional " coordinate" with
a constrain. We hope that  the correspondence of spherical
function on fuzzy sphere  gives a convenient basis and provides a
clear physical picture. The best way to set up such correspondence
is the coherent state technique. In ref\cite{aps}, a coherent
state is constructed with the help of the stereographic
projection. It is not convenient for our purpose. In this paper,
we take the standard method to define the coherent state
\cite{pere}.

\section{ Fuzzy sphere and coherent state}

The algebra ${\cal A}$ of functions on the fuzzy sphere is
defined as a finite dimensional algebra generated by "coordinate"
$x_i, i=1,2,3$ with relations
\begin{equation}
[x_i, x_j]=i\theta\epsilon_{ijk}x_k\quad,\quad
\sum_{i=1}^3x_i^2=r^2
\end{equation}
where the parameter $\theta$ stands for the non-commutativity
and $r$ the radius of the fuzzy sphere. With appropriate value of
$\theta$ ,  this algebra has a finite
number of basis.  The basis  may be represented by the "spherical
harmonic functions $ y^l_m, l\leq N$ with Moyer
product (star product) \cite{ars}
\begin{equation}
Y^{I}_{i}*Y^{J}_{j}=\sum_{K,k}\left[\begin{array}{ccc} I&
J&K\\i&j&k\end{array}\right]
c^{\mbox{k},\alpha}_{IJK}Y^K_k
\end{equation}
We has another metx forhod to realize it as an operator form. The matrixn
form is \cite{hnt}
\begin{equation}
(T_{j,m})_{m_1,m_2}=(-1)^{\frac{p-1}{2}-m_1}\sqrt{2j+1}\left(
\begin{array}{ccc} \frac{p-1}2&j&\frac{p-1}2\\
          -m_1& m&m_2\end{array}\right)
\end{equation}
   The algebra ${\cal A}_N$ has a convenient realization by the
$SU(2)$ Lie algebra. For a $N+1$-dimensional irreducible
representation, the generators of $SU(2)$ satisfy
\begin{equation}
[L_i, L_j]=i\epsilon_{ijk}L_k\quad,\quad
\sum_{i=1}^3L_i^2=\frac{N(N+2)}{4}.
\end{equation}
The basis of  $H$ can be chosen as $|N/2,m>$ with relations
\begin{eqnarray}
L_{\pm}\vert N/2,m\rangle =\sqrt{(N/2\mp m)(N/2\pm m+1)}\vert
N/2,m\rangle \nonumber \\
L_3\vert N/2,m\rangle =m\vert N/2,m\rangle\quad,\quad m=N/2,
N/2-1, \cdots, -N/2
\end{eqnarray}
with $L_{\pm}=(L_1\pm iL_2)/\sqrt{2}$. 

It is well-known that there is
an isomorphism between ${\cal A}_N$ and $SU(2)$ Lie algebra
\begin{equation}
x_i=\lambda_N L_i
\end{equation}
On the $H_N$,  acturally, we will show $Lf=[x\stackrel{*}{,}f]$ on the
function $f$. 
It gives a relation between the radius of fuzzy sphere
and the Carsimir
\begin{equation}
r^2=\frac{\theta^2N(N+2)}{4}.
\end{equation}

   For algebra ${\cal A}_N$, the basis can be written in terms of
the spherical  functions but with non-trivial
multiplication Eq.(1). We will show it explicitly by using the 
generalized coherent states of $SU(2)$ 
\begin{equation}
\vert \;\omega \rangle =T(g)\vert v\rangle\quad,\quad\vert v\rangle\in H_N
\end{equation}
where $T(g)$ is an  element of $SU(2)$ group
\begin{equation}
T(g)=e^{i\alpha L_3}e^{i\beta L_2}e^{i\gamma L_3}.
\end{equation}
These  coherent states satisfy
\begin{equation}
\frac{N+1}{8\pi^2}\int d\,\Omega_3 \vert\omega)(\omega\vert=1
\end{equation}
here $ d\Omega_3$ stands for  3-volume form of the group manifold
 $\Omega(\alpha,\beta,\gamma)$. If taking a gauge $\alpha=\phi,
\beta=\theta,
\gamma=-\phi$ and  $\vert v\rangle =\vert N/2,-N/2\rangle$, then this
gives standard
coherent state \footnote{The explicit form was given in Ref.\cite{pere},
but no relation with
D-function}
\begin{eqnarray}
\vert \omega(\theta,\phi) \rangle &=& \sum_{\mu=-N/2}^{N/2}
D^{N/2}_{\mu,-N/2}(\phi,\theta,-\phi) \vert N/2,\mu\rangle
\end{eqnarray}
Here
\begin{equation}
\begin{array}{rcl}
D^{j}_{\mu,\nu}(\alpha,\beta,\gamma)&=&\displaystyle\sum_k
\frac{(-1)^k\sqrt{(j+\mu)!(j-\mu)!(j+\nu)!(j-\nu)!}}
{k!(j-\nu-k)!(j+\mu-k)!(k-\mu+\nu)!}  \nonumber \\
& &\displaystyle e^{-i\alpha\mu}(cos(\beta/2))^{2j-\nu+\mu-2k}
(sin(\beta/2))^{2k-\mu+\nu}e^{-i\nu\gamma}
\end{array}
\end{equation}
For example  $D^{j}_{0,m}(\phi,\theta,-\phi)=Y^j_m(\theta,\phi)$. 
In this Dirac gauge, the
completeness condition changes into the integral on $S^2$.
\begin{equation}
\frac{N+1}{4\pi}\int \sin(\theta)d\,thetad\,\phi \vert
\omega(\theta,\phi))(\omega(\theta,\phi)\vert =1
\end{equation}
First of all, let $f$ be a function of
$(\theta,\phi)$ and define an operator $\hat{f}$ by
\begin{equation}
\hat{f}= \int
d\,\Omega_2f(\theta,\phi))\vert\omega(\theta,\phi))(\omega(\theta,\phi)\vert
\end{equation}
For a basic function
$y^l_m(\theta,\phi),0\leq l\leq N$, the corresponding operator
$\hat{Y}^l_m$ is
\begin{equation}
\hat{Y}^l_m=\frac{N+1}{4\pi}\int d\,\Omega_2 y^l_m(\theta,\phi)
\vert \omega(\theta,\phi))(\omega(\theta,\phi)\vert.
\end{equation}
Putting  between two vectors in $H_N$ will give the matrix element
of $\hat{Y}^l_m$
\begin{eqnarray}
\left(\hat{Y}^l_m\right)_{\mu,\nu}&=&\langle N/2,\mu\vert
\hat{Y}^l_m\vert N/2,\nu\rangle\nonumber \\
&=& a_{N,l}(-1)^{l-N/2+\mu}\sqrt{N+1}\left( \begin{array}{ccc}
     N/2 & l& N/2\\\nu& m&\mu \end{array}
     \right)
\end{eqnarray}
with \begin{equation}
 a_{N,l}=(-1)^l\sqrt{N+1}\left(\begin{array}{ccc}
      N/2 & l& N/2\\ -N/2& 0& N/2 \end{array}
     \right)
\end{equation}

  In (N+1)-dimensional Hilbert space on which  operator acts, one can
define a
symbol related to the operator through
\begin{equation} \tilde{ F}(\theta,\phi)=(\omega(\theta,\phi)\vert
\hat{f}\vert\omega(\theta,\phi))
\end{equation}
Since the coherent states are not orthogonal, the symbol
$\tilde{F}$ is not equal to $f$. But there must some
relations between them. In ref.\cite{pere}, both $f$ and ${
F}$ are called q-symbol and p-symbol respectively. Somehow it is
related to the (anti)-normal order. To explore such relation, we
put our attention on the symbol  of  typical basis $y^l_m$
\begin{eqnarray}
\tilde{ Y}^l_m(\theta,\phi)&=& (\omega(\theta,\phi)\vert \int
d\,\Omega_2(\theta',\phi')y^l_m(\theta,\phi)\vert\omega(\theta',\phi'))
(\omega(\theta',\phi')\vert \vert \omega(\theta,\phi))\nonumber \\
&=& \sum_{\mu,\nu,J}\langle N/2,-\mu,N/2,\nu\vert J,\nu-\mu\rangle
\langle N/2,N/2,N/2,-N/2\vert J,0\rangle \nonumber \\
& & \langle N/2,\nu,l,m\vert N/2,\mu\rangle
a_{N,l}\frac{2l+1}{\sqrt{4\pi}}(-1)^{\mu+N/2}
D^J_{\nu-\mu,0}(\theta,\phi)\nonumber \\
 &=& a_{N,l}^2\tilde{Y}^l_m(\theta,\phi)
 \end{eqnarray}

With the help of symbol, we can define the "star product" (Moyer
product) of two functions (symbols) to be the symbol of two
operators, namely
\begin{eqnarray}
\lefteqn{\tilde{ Y}^{j_1}_{m_1}*\tilde{
Y}^{j_2}_{m_2}(\theta,\phi)=
(\omega(\theta,\phi)\vert\hat{Y}^{j_1}_{m_1}\hat{Y}^{j_2}_{m_2}\vert
\omega(\theta,\phi))}\nonumber \\
&&=\int d\Omega_2(\theta',\phi')
(\omega(\theta,\phi)\vert\hat{Y}^{j_1}_{m_1}\vert\omega(\theta',\phi'))
(\omega(\theta',\phi')\vert \hat{Y}^{j_2}_{m_2}\vert
\omega(\theta,\phi))\nonumber \\
& &=\sum_{\mu,\nu,J,m}(-1)^{\mu+N/2}a_{N,j_1}a_{N,j_2} \langle
N/2,m,j_1,m_1\vert N/2,\mu\rangle \langle N/2,\nu,j_2,m_2\vert
N/2,m\rangle
  \nonumber \\
& &\;\;\;\;\times \langle N/2,\nu,N/2,-\mu\vert J,\nu-\mu\rangle
 \langle N/2,-N/2,N/2,N/2\vert J,0\rangle
D^J_{\nu-\mu,0}(\theta,\phi)\nonumber \\
&&= \sum_{J,m}\langle j_1,m_1,j_2,m_2\vert J,m\rangle
\sqrt{2J+1}\left\{\begin{array}{ccc}
     j_1& j_2 & J \\N/2&N/2&N/2\end{array}\right\} \nonumber \\
     & &\;\;\;\;\;\;\times
a_{N,j_1}a_{N,j_2}a_{N,J}^{-1}\tilde{Y}^J_m(\theta,\phi).
\end{eqnarray}
where $\{\cdots\}$ stands for $6j$-symbol. Thus, the star product
of two normalized functions ${ Y}^l_m=a^{-1}_{N,l}\tilde{
Y}^l_m$ ( Weyl symbols) is given by
\begin{eqnarray}
{Y}^{j_1}_{m_1}*{
Y}^{j_2}_{m_2}(\theta,\phi)&=&\sum_{J,m}(-1)^{j_2-j_1-m}(2J+1)
a_{N,j_1}a_{N,j_2}a_{N,J}^{-1}\left(\begin{array}{ccc}
     j_1& j_2 & J \\ m_1& m_2& -m \end{array} \right)\nonumber\\
& &\;\times \left\{\begin{array}{ccc}
     j_1& j_2 & J \\N/2&N/2&N/2\end{array}\right\}
{ Y}^J_m(\theta,\phi).
\end{eqnarray}
This is nothing but the relation Eq.(2) appeared in ref.\cite{ars}. A
little difference
is due to the normalization of $6j$-symbol. This provides a
realization of algebra ${\cal A}_N$. It is clear that the symbol (14) is
same as Eq.(3) given in 
Ref\cite{hnt} up to a normalization factor. Therefore, we have found a
correspondence between function and operator realization.  The integral in
function space becomes the 
the trace of operators in Hilbert space, i.e.
\begin{equation}
Tr\Longrightarrow \frac{N+1}{4\pi}\int d\,\Omega_2.
\end{equation}
Comparing with one in non-commutative plan, one can conclude
$\Theta =2/(N+1)$. This was obtained in Ref. \cite{hnt} by
taking the large $N$ limit. Here it is valid for all value of $N$.

  The action of differential operators $L_a,
a=+,-,3$ on symbol is given by 
\begin{equation}
\left(L_a{
Y}^l_m\right)(\theta,\phi)=\sqrt{\frac{N(N+1)(N+2)}{12}}\left[{
Y}^1_a \stackrel{*}{,}{ Y}^l_m\right]
\end{equation}
Using Eq.(21), one can check 
\begin{eqnarray}
r.h.s&&=\sum_j\sqrt{\frac{(2l+1)N(N+1)(N+2)}{12}}
\left\{\begin{array}{ccc}1& l&
j\\N/2&N/2&N/2\end{array}\right\}\nonumber \\
& &\times[<1,a,l,m|j,a+m>-<l,m,1,a|j,a+m>]{\cal Y}^j_{a+m}\nonumber \\
& &=\left\{\begin{array}{l}\sqrt{(l\mp m)(l\pm m+1)}{
Y}^l_{m\pm1} \quad,\quad a=\pm \\
 m{ Y}^l_{m}\quad,\quad a=3 \end{array}
\right.
\end{eqnarray}
It is consistence with the left hand side of Eq.(23). This equation is
same as one appeared in
ref.\cite{ars1} up to a factor. 
Notice that this extra factor on the right hand side of Eq.(23) comes from
the non-commutative 
parameter $\theta^2=(N+1)/2$ and the radius.

\section{Bimodule and Differential operator in moving frame }

Up to now, we have set up the correspondence of the differential
operators. In principle, we are ready to write down a quantum
gauge field theory. However, these three differential operators
are not independent. the fuzzy sphere, since the constrain
$\sum x_i^2=1$,has two independent degree of freedom. Thus, the best
way is to find two independent differential operators. On th usual
sphere, there exist two tangent vector (in moving frame). this can
be done  by introducing the right acting operator on the basis.
Somehow it is not necessary for usual case  \cite{ww}.  This
idea can be generalized into fuzzy sphere. For present situation,
the right acting operators are very  important.  
On fuzzy sphere,  the normal vector relates the spin of frame. 
 Thus,the differential operators along two tangent and normal directions
constitute a right-acting SU(2) group which 
commutative   with the original left-acting SU(2). The totation on the
sphere is a subgroup.  The Hilbert space $H_N$ corresponding to the left
action of algebra ${\cal A}_N$ also provides a $N+1$-dimensional
representation of such SU(2).  Exactly, the 
basis of algebra ${\cal A}_N$ are a bi-module.  
In this section, we will introduce right
acting differential operators.

   Let $J_a, a=+,-,3$  the three right-acting operators in the
moving frame acting on the symbol ${\cal
D}^l_{m,\mu}(\alpha,\beta,\gamma)$ as
\begin{eqnarray}
J_{\pm}{\cal
D}^l_{m,\mu}(\alpha,\beta,\gamma)&=&\sqrt{l\mp\mu)(l\pm\mu+1)}
{\cal D}^l_{m,\mu\pm1}(\alpha,\beta,\gamma)\nonumber \\
J_{\pm}{\cal D}^l_{m,\mu}(\alpha,\beta,\gamma)&=& \mu {\cal
D}^l_{m,\mu}(\alpha,\beta,\gamma)
\end{eqnarray}
Here we  still write the right acting operators at the left of
basis, but the action is different. They act on the second
subscript of symbol ${\cal D}^l_{m,\mu}$ in stead of the first
index. Using the coherent state properties, we can show
\begin{equation}
J_a{\cal D}^l_{m,\mu}=\left[{\cal D}^1_{0,a}\stackrel{*}{,}{\cal
D}^l_{m,\mu}\right]\sqrt{\frac{N(N+1)(N+2)}{12}}
\end{equation}
The proof is very similar with one in left multiplication case. In
fact, this equation provides a method to define a local
coordinates $\hat{x}_{\pm}$ by
\begin{equation}
\hat{x}_{\pm}=\sqrt{\frac{N(N+1)(N+2)}{12}}{\cal D}^1_{0,a},
\end{equation}
then it becomes
\begin{equation}
J_{\pm}{\cal D}^l_{m,\mu}=\left[\hat{x}_{\pm}\stackrel{*}{,}{\cal
D}^l_{m,\mu}\right]
\end{equation}

In Ref.\cite{ccly},  three operatotrs $K_a$ of another SU(2) are found by
using two sets of bosonic 
operators. They are commutative with
the generators $L_a$.  The $K_3$ must correspond to $J_3$ in present paper
describing a rotation freedom of the frame. The other two $K_{\pm}$ act
like $J_{\pm}$ in our notation. They did not discuss it from the point of
view of bimudule. The advantedge of our method is to interplet the $L_a$
and $J_a$ as  the left- right- acting 
operators on the basis of ${\cal A}_N$. The counterpart of $J_a$ on
function space is explicitly constructed. 
We find that this realization is natural and convenient. 
 
\section{ABS construction}

In  soliton theory, the soultion gereating technique has been a
significant tool. Recently, such approach was applied into
non-commutative geometry. In ref.\cite{hkl}the quasi-unitary operator
was introduced and various solutions were obtained for several
theories. The key point is the quasi-unitary operator $S$ which
satisfies $S\bar{S}=1-P, \bar{S}S=1$. This property is closely
related to  fact  the infinite dimensional Hilbert space.  For
fuzzy sphere, however, the dimension of Hilbert space is just
$N+1$. The concept of quasi-unitary operator  must be generalized
if applying ABS method to generate new solutions in finite Hilbert
space.

  On fuzzy sphere, there are two independent degree of freedom.
The method to choose such two coordinates  labelled  by
$\hat{x}_{\pm}$ in the moving frame is given by Eq.(27). Define
two operators (partial isometry) $T$ and $\bar{T}$ by
\begin{equation}
T=\frac1{\sqrt{\hat{x}_-\hat{x}_+}}\hat{x}_-\quad,\quad
\bar{T}=\frac1{\sqrt{\hat{x}_+\hat{x}_-}}\hat{x}_+
\end{equation}
which satisfy
\begin{equation}
T\bar{T}=1-P_{N/2} \quad,\quad \bar{T}T=1-P_{-N/2}
\end{equation}
where $P_i$ is a projecting operator onto i-th dimension in
Hilbert space.  Since $\hat{x}_-$ ($\hat{x}_+$) has a kernel
$|N/2,-N/2>$ ($|N/2,N/2>$), the $T$ ($\bar{T}$ has also same
kernel as $\hat{x}_-$ ($\hat{x}_+$). This is ensured by  choosing
properly the order in the denominator of $T$ and  $\bar{T}$. Such
partial isometry operators will play an important role in
constructing the new soliton-like solutions. In ref.\cite{ccly},
A similar result is obtained. But it was given in terms of operators. In
present,  all things 
are  based on the partial isometry releted to two tangent vectors. It can
considered as the 
function realization. Using the same method, one can analyzes CP(n) model
on fuzzy sphere.

\section{BPS solitons}

  Consider a complex scalar field theory on fuzzy sphere. The
action reads
\begin{equation}
S=\int d\;\Omega_2D_a\Phi D^a\Phi
\end{equation}
where $D_a\Phi=J_a\Phi-i\Phi*A_a$. The equation of motion is
\begin{equation}
D_a\Phi=0
\end{equation}
Since the gauge field $A_a$ has no kinetic term, we can get
\begin{equation}
A_a=-i\Phi^{\dag}*J_a\Phi \end{equation} Thus the equation of
motion can be written as
\begin{equation}
D_a\Phi=J_a\Phi-i\Phi*(-i\Phi^{\dag}*J_a\Phi=(1-\Phi*\Phi^{\dag})*J_a\Phi=0
\end{equation}
It is clear that the system  has a trivial solution $\Phi=const.$
and $A_a=0$. Now, we try to give other non-trivial solutions.
Assume $\Phi=T^n, n\leq N$. Then one check that it is a new
solution
\begin{equation}
D_aT^n=(1-T^n\bar{T}^n)*J_aT^n=P_{N,N-n}J_aT^n=0
\end{equation}
where $ P_{N,N-n}=P_N+P_{N-1}+\cdots+P_{N-n+1}$.

 In the above discussion, we did not consider the affection of
potential. It is quite easy to generalize into including potential
case. Suppose the potential to be form
$V(\Phi^{\dag}\Phi-|\Phi_0|^2)$ which has an extrema at $\Phi=0$
and a local minimum at $\Phi=\Phi_0$. Due to the appearance of
potential, the equation of motion should be modified. Namely, the
right hand side of Eq.(30) to be
$V'(\Phi^{\dag}\Phi-|\Phi_0|^2)\Phi^{\dag}$ instead of zero.
Putting $\Phi=T^n$ and suing the property of $T$, one can show the
added term in equation of motion vanishing. So, it keeps the
equation of motion unchanged.

   It is worthy to point out that these solutions are the
eigen-state of $J_0$ with value $(-2n)$.  Similarly, one can also
choose $\Phi=\bar{T}^n$ which will give another kind of solutions
with the eigen-value $(2n)$ of $J_0$. A similar results were also
obtained in ref\cite{ccly} for $CP(N)$ model on fuzzy sphere.

\section{Flux-like solution}

 In this section, we will discuss another kind of solution in gauge field
theory with a scalar
field $\Phi$. The Lagrangian is taking the form
\begin{equation}
{\cal L}=\frac{-1}{g}\int d\,\Omega_2(\frac14F_{ab}F^{ab}+D_a\Phi
D^a\Phi)
\end{equation}
where $\Phi$ takes the adjoint representation, i.e.
\begin{equation}
D_a=J_a\Phi+[A_a\stackrel{*}{,}\Phi]
\end{equation}
The equation of motion reads
\begin{eqnarray}
 \left[D_a, \left[D^a,D^b\right]\;\right]+
\left[\Phi, \left[\Phi, D^b\right]\;\right]=0\nonumber \\
\left[D_a, \left[D^a,\Phi\right]\;\right]=0 \label{eom2}
\end{eqnarray}
It is easy to check that the Eq.(\ref{eom2}) has a trivial
solution $\Phi=const.$ and $A_a=0$. Let us consider another
solution $\Phi=T^n\bar{T}^n=1-P_{N,N-n}$. First, we need show it
to be a solution.

  Using the explicit expression, one can find
\begin{equation}
[D_a,
\Phi]=T^nJ_a\bar{T}^nT^n\bar{T}^n-T^n\bar{T}^nT^nJ_a\bar{T}^n=0
\end{equation}
and the first equation of Eq.(\ref{eom2}) is also valid.  The
gauge field strength is given by
\begin{eqnarray}
F_{+,-}&=&[D_+,D_-]-2D_3-([J_+,J_-]-2J_3)=n(N+1)P_N\nonumber\\
F_{3,\pm}&=&0
\end{eqnarray}

\section{D-brane on fuzzy sphere}

  Let us discuss the tachyon condensation on non-BPS D-brane on fuzzy
sphere. As done in non-commutative $R^n$ space, we chose the background
field $B$ which is not constant in our case. On any non-BPS D-brane there
exists a tachyon field $\phi$ ( do not confuse with the partial isometry
operator in above section) and a gauge field $A_a$. The effective action
of a non-BPS D2-brane can bewritten as the DBI form \cite{garo}.
In our case it reads
\begin{equation}
S=\frac{\sqrt{2}}{\sqrt{\alpha'}G_s}\int d\,t
Tr\left[V(\phi)\sqrt{det(G+2\pi\alpha'(F+B))}\right]+O\left(D_a\phi,
D_aF\right)
\end{equation}
where $D_aT, D_aF$ denote the covariant derivative of the tachyon and
the gauge field strength on fuzzy sphere; the last term means including
the higher derivative of $T$ and $F$. As argued in many cases, the tachyon
condensation does not depend on the detailed form  of the last term. This
property is still valid in fuzzy sphere case.  The  Tachyon  potential 
factor appeared in the front of DBI form is followed from Sen's
conjecture. It contains a local maximum $T=0$ and  local mimimum
$\phi=\phi_0e^{i\theta}$. 

  On the non-BPS D-brane, both  tachyon  and gauge fields take the adjoint
representation of the gauge group. Based on the correspondence between
Moyer product and operator product, the derivative could be considered as
the operator on fuzzy sphere. First, we want to solve the equation of
motion of the tachyon field. The general form  is
\begin{equation}
D_aD^a\phi+\cdots=\frac{\partial V(T)}{\partial T}
\end{equation}
Here $\cdots$ stands for terms related to higher derivative of $T$.
The solutions of this equation can be represented by the projection 
operator on fuzzy sphere $\phi=\phi_0 (1-P_{N/2,n})$. The  projection  on
fuzzy
sphere consists of the partial isometry operator (88). Since $1-P_{N/2,n}$
is also a projection, we have
$$\frac{\partial V(T)}{\partial T}|_{T=T_0(1-P_{N/2,n})}=
\frac{\partial V(T_0)}{\partial T_0}(1-P_{N/2,n})=0.               
$$
On the other hand, one can check that such tachyon solutions satisfy a
simple equation $D+_\phi=D_+\phi=0$.  Thus this fact ensures the whole
equation
of motion even if we do not know the explicit form  of second term in
Eq.(11).  Next we examine the gauge field. The equation of motion of gauge
field is $D_aF^{ab}=0$. This is automatically satisfied if the strength is
proportional to a projection. For our case, the detail calculation shows
the gauge field strength to be  $F_{+-}=n(N-n)P_{N/2}$. The mass of this
excitation is
\begin{equation}
M=\frac{\sqrt{2}}{\sqrt{\alpha'}G_s}
Tr\left[V(\phi_0)(1-P_{N/2,n})\sqrt{det(G+2\pi\alpha'(F+B))}\right]
\end{equation} 

The last task of this section is to investigate the tachyon condensation
on brane-antibrane systems. It is much more complicated that non-BPS
D-brane. Of reasons is the tachyon becoming into complex and belonging to
a bi-module. The actions of  operators  from the left and right of
bi-module are different \cite{ttu,kmm,kmm1}.

The effective action of a $D2-\bar{D}2$ with two gauge fields has been
computed through boundary string field theory \cite{ttu,kmt}.
It was applied to non-commutative tori in \cite{bkmt,mar}. For our case
the effective action takes
\begin{eqnarray}
S& =&\frac1{\sqrt{\alpha'}G_s}\int d\,t
Tr_1\left[V^{(1)}(\phi\bar{\phi})\sqrt{det(G+2\pi\alpha'(F^{(1)}+B))}\right]
\nonumber \\
& &
\&frac1{\sqrt{\alpha'}G_s}\int d\,t
Tr_2\left[V^{(2)}(\bar{\phi}\phi)\sqrt{det(G+2\pi\alpha'(F^{(2)}+B))}\right]
\nonumber \\
& & 
+O\left(D_a\phi,
D_aF^+, D_aF^-\right)
\end{eqnarray}
where $O(x)$ denotes the derivative of tachyon and two gauge fields. The
tachyon potentials $V^{(i)}$ are assumed to  be staionary at $T_0$. From
the action, the equations of motion are
\begin{eqnarray}
D_a\phi=\bar{\phi}\frac{\partial
V^{(1)}(x)}{\partial(x)}|_{x=\phi\bar{\phi}}
\nonumber \\
D_a \bar{\phi}=\phi\frac{\partial V^{(2)}(x)}{\partial(x)}|_
{x=\bar{\phi}\phi}         
\nonumber \\
D_aF^{(i)}=0  
\end{eqnarray}
Choose the following solution 
\begin{equation}
\phi=\phi_0T^n\quad,\quad \bar{\phi}=\phi_0\bar{T}^n
\end{equation}
one can check that they give the stationary points of the tachyon
potential. A detailed calculation shows these solutions satisfying the
equation of motion of tachyon. For gauge field, we choose
\begin{equation}
A_a^{(1)}=0\quad,\quad A_a^{(2)}=T^n\hat{j}_a\hat{T}^n-\hat{j}_a
\end{equation}
The proof of gauge field satsifying
equation of motion is straightforward.  
The gauge field strenghth is $F^{(2)}_{+,-}=n(N-n)P_{N/2}$.

\section{Discussion}
In this paper, we propose that the Hilbert space of algebra ${\cal A}$ is
a bo-module. The operators acting on the bi-module are considered as the
differential operators in both fixed frame and moving frame. Based on the 
two tangent vectors on fuzzy sphere, we  carried out  the ABS construction
 on fuzzy sphere and applied into the soliton and flux solutions of gauge
field theory.  The application to D-brane systems and the mass psectrum
are discussed. 

  In Ref. \cite{ccly}, the bosonic realization of SU(2) algebra are used.
Since the $L_a$ is not a proper derivative to expose topologically
nontrivial field configurations, Chan et al proposed another SU(2) $K_a$.
The operator differential equation looks like Eq.(26) (Eq.(2.24) in
ref.\cite{ccly}). The BPS solution obtained in ref.\cite{ccly} must equal
to Eq.(35). Since the BPS solitons are given interms of partial isometry,
it is not difficult to move a way from the origin by shifting the
parameters in symbols. Thess new solutions are similar to the $W^{\pm}_k$
(Eq.(4.12) in ref.\cite{ccly}).  Thus, the results in section 5 partially
recover those in ref.\cite{ccyl}.

  It would be desirable to extent the analysis to other fuzzy spheres such
as $S^3$ and $S^4$. For $S^3$, it can be considered as a coset of
$SO(4)/SO(3)$. Using the similar method, one may construct the ABS
operators and investigate the related noncommutative Yang-Mills theory.

{\bf Acknowledgment}

We are grateful to Y.S. Wu for useful discussions and Miao Li for
discussions and the hospitality during the Summer School on Strings 
from July 16-27. B.Y. Hou also likes to thank Martinec for discussions.
We also acknowledge the NSFC for support.

 \end{document}